\documentclass[reprint,
 amsmath,amssymb,
 aps,
 pra,
]{revtex4-1}

\usepackage{graphicx}
\usepackage{dcolumn}
\usepackage{bm}

\begin{document}

\preprint{APS/123-QED}

\title{Spontaneous energy-barrier formation in an entropy-driven glassy dynamics}

\author{Chiara Cammarota}
 \email{chiara.cammarota@roma1.infn.it}
\affiliation{
 "Sapienza", University of Rome\\
 P.le Aldo Moro 2, I-00185 Rome, Italy
}
\author{Enzo Marinari}
 \email{enzo.marinari@uniroma1.it}
\affiliation{
 "Sapienza", University of Rome and IPCF-CNR, UOS Rome \\ P.le Aldo Moro 2, I-00185 Rome, Italy\\
}

\date{\today}

\begin{abstract}
The description of activated relaxation of glassy systems in the multidimensional configurational space is a long-standing open problem. 
We develop a phenomenological description of the out-of-equilibrium dynamics of a model with a rough potential energy landscape and we analyse it both numerically and analytically. The model provides an example of dynamics where typical relaxation channels go over finite potential energy barriers despite the presence of less-energy-demanding escaping paths in configurational space; we expect this phenomenon to be also relevant in the thermally activated regime of realistic models of glass-formers. In this case, we found that typical dynamical paths episodically reach an high fixed threshold energy unexpectedly giving rise to a simple thermally activated aging phenomenology. In order to unveil this peculiar aging behavior we introduce 
a novel description of the dynamics in terms of spontaneously emerging dynamical basins.

\begin{description}
\item[PACS numbers]
May be entered using the \verb+\pacs{#1}+ command.
\end{description}
\end{abstract}

\pacs{Valid PACS appear here}
\maketitle


{\it Introduction.}{\bf--}Glass formation manifests itself as a dynamical phenomenon. Below a crossover temperature $T_g$, supercooled systems (characterized by aperiodic disordered patterns in the metastable liquid phase) run out of equilibrium on the timescale of experimental observations~\cite{edanna96}. 
When approaching from above the temperature of the glass formation crossover, the growth of the relaxation timescale is faster than an Arrhenius behavior, $\tau\sim\exp(\Delta/T)$, where $\Delta$ is a fixed activation barrier and $T$ is the temperature. 
This observation originally suggested that the progressively slower relaxation dynamics would be the result of activated processes involving increasing activation barriers $\Delta(T)$ when $T$ decreases~\cite{goldst69}.
\\ 
Providing a quantitative description of glassy dynamics in terms of activated processes is still an open problem. A long theoretical effort started with the definition of simple but interesting models for thermal activation, called {\it trap models}~\cite{boucha92,boudea95}. New results~\cite{bencer06,beboga02,gayrar10,gayrar14,foisns01} recently confirmed the broad interest of the trap model paradigm for the description of realistic activated dynamics and inspired the present work.
A number of alternative pictures have been proposed in the literature, ranging from a microscopic dynamical theory of non activated relaxation~\cite{gotsjo92}, called Mode Coupling Theory, and other attempts towards the formulation of a purely dynamical picture free from any landscape influence~\cite{freand84,chagar10}, to
a peculiar thermodynamic theory of the slow dynamics, the Random First Order Transition (RFOT) theory~\cite{kithwo89,cavagn09,berbir11,parzam10,chkpuz14}, whose main ingredient is the rough (complex) structure of the potential energy landscape (PEL) in a multidimensional configurational space. 
\\
In these approaches, the large dimensionality of the configurational space can play different roles both in the equilibrium and out-of-equilibrium dynamics.
On the one side, it is possible to show that the activated relaxation through a barrier of high $\Delta$ in any direction of a large dimensional space should occur on average on a timescale smaller than the expected Arrhenius $\tau\sim\exp(\Delta/T)$~\cite{kurcha11}. The entropy of possible escaping directions produces as an effect the lowering of the activation barrier.
On the other side, 
for non activated processes, 
the dynamics is slowed down by the presence of an overwhelming number of directions in the configurational space that do not allow the system to relax till one of the rare escaping channels is found (entropic bottleneck effect)\cite{kurlal96,tonbir07,olkrre11}.
\\
These two mechanisms have been investigated separately, respectively within the thermodynamic-influenced RFOT picture and the non-activated purely dynamical approach to the problem of glass formation. However
it is reasonable that for the relaxation dynamics in complex landscapes 
we could deal with a distribution of dynamical paths that includes at the same time frequent relaxation channels involving the hopping of large barriers and rare paths characterized by the smallest possible barriers.
In this case the large dimensionality of the configurational space can play the two previously discussed roles at the same time.
In this Letter we analyze the problem of activated dynamics in a multi-dimensional configurational space from this particular perspective. 
We mainly focus on the out-of-equilibrium and aging properties of the dynamics.
We discuss a model in which an entropy of paths (originated by the frequency of possible relaxation processes) and the activation energy compete during the dynamics, selecting the fastest relaxation channels among the more frequent but activated ones, despite the presence of non-activated alternatives.
In this case, the contribution of the entropy of paths surprisingly gives rise, in a long-time coarse-grained description, to a genuine out-of-equilibrium relaxation behaviour typical of standard activated dynamics (trap models).

{\it The model.}{\bf--}Our system of interest can be in $M$ configurations $i$, with energy $E_i<0$ given by independent identically distributed random variables extracted from an exponential distribution $\rho(E)=\lambda\exp(\lambda E)\hspace{0.6mm} \theta(-E)$. We introduce a Metropolis dynamics with non-zero transition rates from any level $i$ to any other level $j$ given by 
\begin{equation}
w_{i,j}=
\left\{
\begin{array}{lr}
1 & \text{if} \; E_i\geq E_j \\
\exp[-\beta(E_j-E_i)] & \text{if} \; E_i< E_j
\end{array}
\right.
\ .
\end{equation}
\\
This model is equivalent to the one defined by Mezard and Barrat~\cite{barmez95,bertin10} and called the {\it step model}. 
They used a Glauber dynamics 
instead of our Metropolis rule. 
\\
Step models were introduced in antithesis to preexisting trap models.
The latter were first used by Bouchaud~\cite{boucha92} as the paradigm of low temperature activated dynamics in a multi-minima PEL. In the trap case one considers the same exponential distribution $\rho(E)$, 
but each configuration $i$ occupies the bottom of a trap ideally surrounded by energy barriers $\Delta_i=-E_i>0$ in any direction. The transition rates read $w_{i,j}=\frac{1}{M} \exp(-\beta \Delta_i)$
and the time for escaping from each configuration $\tau_i\propto 1/w_{i,j}\propto\exp(\beta \Delta_i)$ simply follows an Arrhenius law. 
At variance, step models were designed without the explicit introduction of potential energy barriers to study the influence of the entropy of possible relaxation paths in the context of a non-activated slow dynamics of glassy systems.\\
Here, we focus on step models to study the contribution of the entropy of paths together with activated relaxation, as it should be for deeply supercooled liquids.
Each dynamical move is in fact decided by the competition between a possible energy increase, favoured by an exponentially large number of high-energy configurations and hampered by a low acceptance rate, and an exponentially small number of descending paths with high acceptance rate. The entropic and thermal dynamical drives are quantified respectively by $\lambda$ and $\beta=1/T$.
The study of step models is also motivated by the recent interest~\cite{beboga02,gayrar14} in a possible generalization of the trap-like aging phenomenology to models with more realistic dynamical rules, as for example the Metropolis one.\\
First studies of the step model~\cite{barmez95} (and recently~\cite{bertin10}) found that its low temperature slow relaxation only follows non-activated dynamical paths.
A first evidence of an activated dynamical regime at intermediate temperatures was presented in Ref.\cite{bertin03}.
However, even in the regime where thermal activation sets in, typical sojourn times in single configurations were found to be always determined by entropic mechanisms~\cite{bertin03}. An explicit thermal activation was re-introduced in the dynamics as the best means to recover the low temperature activation features of more realistic systems~\cite{bertin03}.
\\
Here we ignore the fully entropic low temperature range, and we avoid the explicit introduction of thermal activation.
We focus instead on the intermediate temperature regime to provide 
a paradigm and an explanation for the {\it spontaneous} formation of activation barriers during the dynamics, as a pure result of the competition between entropic and energetic dynamical drives. 
As it was suggested by previous intriguing results~\cite{bertin03}, these barriers do not directly give rise to usual thermal activation.
Yet, in our study we are able to show that in this entropy-energy ruled dynamical regime an effective trap-like out-of-equilibrium phenomenology typical of models with genuine thermal activation unexpectedly emerges.

\begin{figure}[ht]
\hspace{-0.8cm}
\includegraphics[height=0.4\textwidth, angle=0]{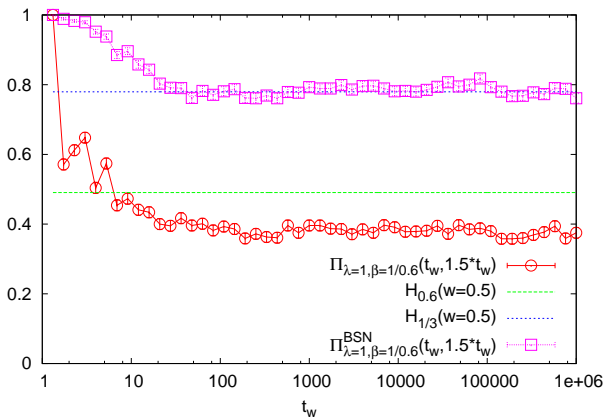}
\vspace{-1.2cm}
\caption{$\Pi_{\lambda=1,\beta=1/0.6}(t_w,t_w+0.5t_w)$ does not converge, for large $t_w$, to $H_{0.6}(w=0.5)$ 
Instead, the probability of not to changing basin $\Pi^{BSN}_{\lambda=1,\beta=1/0.6}(t_w,t_w+0.5t_w)$ converge in the large $t_w$ limit to $H_{2-1/0.6}(w=0.5)$.\label{PItw}} \end{figure}

{\it Two dynamical regimes in step models.}{\bf--}
We study the $M=\infty$ case, where the system can access an infinite number of energy levels. At each step, we start from a configuration with energy $E_i$ and we propose as a trial energy $E_j$, a random variable extracted from the distribution $\rho(E)$. If the Metropolis move is accepted, with rate $w_{i,j}$, $E_j$ becomes the new 
\begin{figure}[ht]
\hspace{-0.8cm}
\includegraphics[height=0.4\textwidth, angle=0]{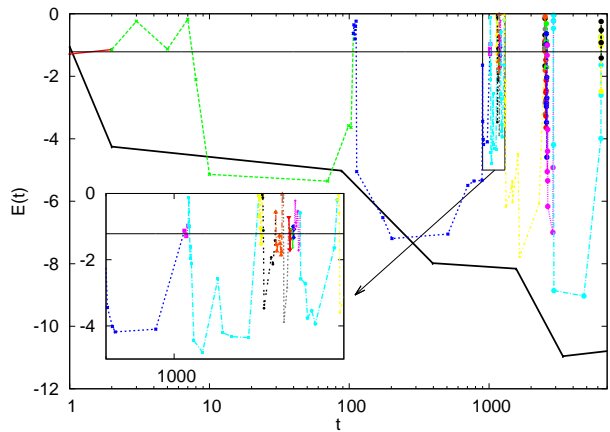}
\vspace{-1cm}
\caption{Two time series (and a zoom of one of them in the inset) of the energy explored by step models for $\rho(E)$ with $\lambda=1$. The continuous line is for $T=0.25$. Colored broken lines are for $T=0.75$. The orizontal line marks the level of the threshold energy $E_{th}$ for $T=0.75$, when it can be defined. Different colors are for different basins, according to the definition in the text.\label{Evst}} 
\end{figure}
energy of the system, otherwise the energy $E_j$ is disregarded, and the old energy $E_i$ is kept. Time always increases by one unit.\\
Numerical simulations show that the model switches, when $T$ changes, between two completely different behaviors.
For low or zero temperatures we observe the entropy-ruled regime studied by Barrat and Mezard~\cite{barmez95,bertin10}. The level of the explored energies continuously decreases as $t$ increases (see the continuous thick line in Fig.\ref{Evst}). For low energies the trapping time in each configuration becomes larger and larger 
since moves that increase the energy are rarely accepted, and moves that lower the energy are only seldom proposed.\\
At intermediate temperatures something completely different occurs. As shown by the broken line in Fig.\ref{Evst} (the dash-color code of its segments will be discussed later), 
the energy 
episodically returns to high values.
This phenomenon signals the spontaneous emergence of effective barriers that the system has to overcome during the dynamics.
Dynamics is fast at high energies but, at large $t$, deeper and deeper configurations are progressively explored arresting the dynamics for increasingly long times. \\
Bertin found first indirect evidences~\cite{bertin03} of a thermally activated behavior in this second regime by observing the large $t/t_w$ power law decay of correlation functions 
between $t_w$ and $t_w+t$.
However, a clear discrepancy emerges comparing the numerical results of the probability to do not change configuration between $t_w$ and $t+t_w$, $\Pi_{\lambda,\beta}(t_w,t_w+t)$, and its large $t_w$ limit  
$C_{\lambda,\beta}(w)=\lim_{t_w\rightarrow\infty;t/t_w=w}\Pi_{\lambda,\beta}(t_w,t_w+t)$ 
in the step model to the theoretical expectations for trap models~\cite{boucha92,boudea95,bencer06,beboga02}.
In trap models for any choice of parameters $\lambda$, $\beta$~\cite{boudea95,bencer06} one has
\begin{equation}
C_{\lambda,\beta}(w)=H_{\lambda/\beta}(w) \ ,
\end{equation}
where 
\begin{equation}
H_{x}(w) \equiv \frac{\sin(\pi x)}{\pi}\int_w^{\infty}du\frac{1}{(1+u)u^x}\ .
\end{equation}
As we show in Fig.\ref{PItw}, $C_{\lambda,\beta}(w)$ for step models has a finite and 
well defined value but it is different from $H_{\lambda/\beta}(w)$.\\
The main feature that determines the aging behavior of the trap models~\cite{boucha92} is  
the fat tail power law distribution of trapping times $p(\tau)\sim \tau^{-(1+\mu)}$ (with $0<\mu<1$) which does not change with the observation time $t$ and whose exponent $\mu=\lambda/\beta$ controls the parameter of $H_x(w)$: $x=\mu$.
In step models, the Metropolis dynamical rule gives rise to a typical distribution of trapping times which ages during the dynamics~\cite{barmez95,bertin10,bertin03} but which at $t\gg\tau$ also behaves like $p(\tau)\sim\tau^{1+\mu}$, with $\mu=2-\beta/\lambda$.
Hence one could also expect an effective trap model behavior described by an $H_x(w)$ with parameter $x=2-\beta/\lambda$. 
Fig.\ref{PItw} shows that numerical data are in contradiction with this expectation, too. Moreover, if we try to extract empirically the effective parameter $x$ assuming that $C_{\lambda,\beta}(w)=H_x(w)$ we find different results for different values of $w$. In conclusion, the trap paradigm apparently does not apply to the $C_{\lambda,\beta}(w)$ obtained for step models.\\
This analysis confirms the deep difference between step and trap models, besides the emergence of dynamical effective barriers in the first case.
As a confirmation of this result we give an estimate of trapping times in the step model.
For every energy level $E$ the probability of going upward $P_{\uparrow}(E)$ and downward $P_{\downarrow}(E)$ are 
\begin{eqnarray}
P_{\uparrow}(E)=\int_E^0 dE' \rho(E') \exp(-\beta (E'-E))=\nonumber \\
=\frac{\lambda}{\beta-\lambda}(\exp(\lambda E)-\exp(\beta E)) \ ,
\end{eqnarray}
and
\begin{equation}
P_{\downarrow}(E)=\int_{-\infty}^E dE' \rho(E')=\exp(\lambda E) \ .
\end{equation}
Hence in the low temperature range, where $\beta>\lambda$, at the leading order the trapping time in each configuration $\tau(E)= \frac{1}{p_{UP}(E)+p_{DO}(E)}\sim \exp(-\lambda E)$ does not depend on the temperature, as it would for thermal activated relaxation channels.

{\it A coarse-grained description of the Metropolis dynamics.}{\bf--}
To make a step forward in the study of this anomalous activated regime, also inspired by Refs.\cite{derebo03,dolheu03,heuer008}, we analyze the dynamics of the model from a different point of view.
We start by focusing on the large {\it dynamical basins} that periodically appear in the microscopic time sequence of the explored energy levels.
These basins stem from the fact that for high $E$ one has $p_{\uparrow}(E)<p_{\downarrow}(E)$ while, for $0.5<\lambda/\beta<1$ (which coincides with Bertin's thermal activated regime~\cite{bertin03,bertin12}), and $E<E_{th}\equiv\dfrac{1}{\beta-\lambda}\ln\left(\dfrac{2\lambda-\beta}{\lambda}\right)$, the opposite inequality holds: $p_{\uparrow}(E)>p_{\downarrow}(E)$.
This means that if the dynamics goes below $E_{th}$ (represented in Fig.\ref{Evst} by the thin orizontal line) it will be driven back to larger energy levels $E>E_{th}$ before falling again deep down in the PEL. This threshold energy plays the role of a dynamical potential energy barrier self-generated by the competition between rare downward descents and exponentially suppressed Metropolis transitions towards higher energies. 
We will therefore use $E_{th}$ to define spontaneously forming large dynamical basins.
Each time $E$ increases and crosses $E_{th}$ we say that the system changes basin (and we change color and line-style in Fig.\ref{Evst}).
 \\
Within this construction in terms of basins we can now define a new correlation function $\Pi^{BSN}_{\lambda,\beta}(t_w,t_w+t)$ as the probability of not changing basin between $t_w$ and $t_w+t$ and study its large $t_w$ limit, $C^{BSN}_{\lambda,\beta}(t/t_w=w)$.
We can also study the probability distribution function of the trapping time in each basin $p(\tau^{BSN})$, for $t\gg\tau^{BSN}$.\\
Not surprisingly, the trapping times distribution changes when we consider basins instead of configurations, \begin{figure}[ht]
\hspace{-0.8cm}
\includegraphics[height=0.4\textwidth, angle=0]{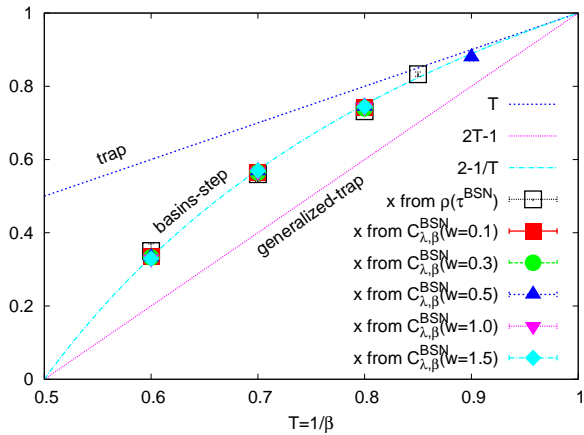}
\vspace{-1.cm}
\caption{The aging parameter $x$ (obtained from the exponent of the power-law tail of $\rho(\tau^{BSN})$ and from the solution of the $H_{x}(w)=C^{BSN}_{\lambda,\beta}(w)$ inverse problem) is reported for a step model with $\lambda=1$ as a function of the temperature. We also show the expectation for $x$ in simple trap models, generalized trap models, and step models. \label{xT}} 
\end{figure}
but the power law behavior for large $\tau$ is described by the same exponent $1+2-\beta/\lambda$ as before\footnote{The tail exponent of distribution of trapping time in  basins has been also recently obtained for a generalization of the step model in the context of the on-off intermittency problem~\cite{bertin12}. 
} (see Fig.\ref{xT}).\\
In practice the dynamical behavior in terms of a basins description of a step model with $\beta\in[\lambda,2\lambda)$ exactly maps onto usual aging of trap models with $\beta_T\in[\lambda,\infty)$ if their aging parameters $x$ correspond: $\lambda/\beta_T=2-\beta/\lambda$.\\
In Fig.\ref{xT} we compare the parameter $x$ with the expectation coming from a naive correspondence to trap models, $x_T=\lambda/\beta$, 
and with the aging parameter $x_{GT}$ for generalized trap models where transition rates depend at each step on initial and final configurations with the same weight: $w_{i,j}(a)=\exp(-\beta a E_i+\beta(1-a)E_j)$~\cite{rimabo00,rimabo01,monthu03,bencer06} with $a=0.5$. 
In this case one can show~\cite{gayrar10} that $x_{GT}=(\lambda/\beta-a)/(1-a)$ (see~\cite{cammar14} for a discussion of the physical meaning of this result), hence $x_{GT}=2\lambda/\beta-1$ for $a=0.5$.
The parameter $x$ obtained in step models differs only by a factor $\lambda/\beta$ from $x_{GT}$. Once more, this result highlights the difference between the entropy-ruled activation of step models (where trapping time is $\tau\sim\exp(-\lambda E)$) and the thermal activation of trap models (where $\tau\sim\exp(-\beta E)$) and excludes any qualitative difference in the long-time dynamical behavior.

{\it Discussion and conclusion.}{\bf--}We have studied
the out-of-equilibrium dynamics of a simple model with rough PEL called step model. In step models potential energy barriers are not introduced {\it a priori}, as opposed to trap models where explicit thermal activated relaxation mechanisms occur. 
Nevertheless, in an intermediate temperature regime $0.5<\lambda/\beta<1$, a non-trivial interaction between entropy and energy of typical relaxation paths takes place. As a result, the dynamics shows the presence of deep basins separated by barriers that form spontaneously, reaching a threshold energy $E_{th}$.
$E_{th}$ directly stems from the entropy trapping mechanism: the larger the rate $\lambda$ of the energy distribution, the less escaping directions with low energy activation cost are available, and the larger is the resulting threshold energy $E_{th}$.
Moreover the trapping times in configurations, $\tau$, and basins, $\tau^{BSN}$, are entirely determined by the entropy of escaping directions.
In spite of that, the typical phenomenology of the genuine trap paradigm, classically referred to thermal activation, can be recognized in the aging behavior of this model provided a description of the dynamics in terms of the spontaneously-formed basins is adopted.\\
The trap-like aging behavior is hence extended by this study to an entropy-energy ruled dynamics like the one of step models.
The introduced coarse-grained basin description of the dynamics plays a major role in this analysis. It does not change the power-law tail behavior of the distribution function of the typical trapping times, a necessary condition for trap model-like aging. On the other hand, it takes advantage of the existence of a reference threshold energy to transform the step model dynamics into a renewal process~\cite{boudea95}: each time the threshold energy is reached the dynamics looses memory of the past hence and subsequent basins are always mutually independent. This was the missing link for establishing the trap model aging phenomenology in the step model. 
\\
The step model provides an entropy-energy ruled relaxation process that could potentially characterize slow dynamics in more realistic glassy systems.
Also in these cases, relaxation in the multi-dimensional configurational space could be controlled by a competition between the energy of the typical barriers and the entropy of the possible dynamical paths. The same competition could determine a specific threshold energy for relaxation processes (the importance of a threshold energy for glassy dynamics was firstly pointed out in Ref.\cite{cugkur93}), transforming generic out-of-equilibrium dynamics into a renewal process.
In this context, the basin description that we have introduced would be a fundamental analytical tool to recognize the possible presence of an underlying trap-model-like aging behavior.

\begin{acknowledgments}
We are grateful to Gerard Ben Arous, Eric Bertin, Giulio Biroli, Jean-Philippe Bouchaud, Jorge Kurchan, and Gilles Tarjus for interesting remarks and discussions.
We acknowledge support from the ERC grant CRIPHERASY (no. 247328).
\end{acknowledgments}

\bibliographystyle{unsrt}
\bibliography{bib}

\end{document}